\documentclass[11pt]{article}
\usepackage[numbers, square]{natbib}
\usepackage{graphicx}
\usepackage[margin=10pt,font=small,labelfont=bf,labelsep=period]{caption}

\usepackage{latexsym}

\def\bf{\textbf}

\def\be{\begin{equation}}
\def\ee{\end{equation}}
\def\bea{\begin{eqnarray}}
\def\eea{\end{eqnarray}}

\def\case#1/#2{\textstyle\frac{#1}{#2}}

\begin{document}

\vspace{.7in}

\begin{center}
\Large\bf {Mach's Principle: A Response to Mashhoon and Wesson's
Paper arXiv: 1106.6036}

\vspace{.2in} \normalsize \large{\textbf{Julian
Barbour}}\,\footnote{Email: Julian.Barbour@physics.ox.ac.uk,
julian@platonia.com.}

\end{center}

\normalsize

\vspace{.2in}

{\textbf{Abstract.} In their recent ``Mach's principle and
higher-dimensional dynamics'', Mashhoon and Wesson argue that Mach's principle
is not properly incorporated into general relativity and that in Einstein's
theory ``the origin of inertia remains essentially the same as in Newtonian
physics.'' While it is true that the motion of a single test particle in a
Newtonian inertial frame of reference appears essentially the same as in an
Einsteinian local inertial frame, this misses the point. The issue
is not what motion looks like in an inertial frame of reference but what is the
origin of the inertial frame. Unlike Newtonian dynamics, general relativity does
implement Mach's principle when considered from this correctly formulated point
of view.

\section{Introduction}

In arXiv: 1106.6036 \cite{Mashhoon2011}, Mashhoon and Wesson argue that Mach's
principle is not properly incorporated into general relativity (GR) because in
it ``the origin of inertia remains essentially the same as in
Newtonian physics'': the motion of a single
test particle in a Newtonian inertial frame of reference appears essentially the
same as in an Einsteinian local inertial frame. But Mach's concern was not
the description of the motion of a single particle in an inertial frame
of reference but \emph{the origin of the inertial frame}. I shall show
in this note that in Newtonian theory there is no mechanism that determines
inertial frames whereas there is in GR. 

The key to establishing this difference between the two theories is the
identification of a precise defect in Newtonian theory (Sec.~2). In Sec.~3, I
show how this defect can be eliminated in a simple Machian model of
nonrelativistic particles by a process called \emph{best matching}. This is
already sufficient to show how Mach's principle should be implemented, and
I therefore merely indicate briefly in Sec~4 how the same mechanism of best
matching represents the core of GR when treated as a dynamical
theory. This shows that GR is Machian in a way that Mach would
have approved. In Sec.~5, I briefly comment on a response to my note
by Mashhoon and Wesson.

\section{The Defect of Newtonian Dynamics}

Mach \cite{Mach1883} argued that in dynamics position must be defined by
observable distances from other particles and not by absolute space or
an inertial frame of reference. However, Mach was an intuitve thinker and,
apart from making it clear that the complete universe must be involved in the
determination of inertial motion,\footnote{A typical comment is: ``When,
accordingly, we say that a body preserves unchanged its direction \emph{in
space}, our assertion is nothing more or less than an abbreviated reference to
\emph{the entire universe} (\cite{Mach1960}, p. 286, Mach's emphasis.)}
did
not give a precise mathematical criterion that would allow one to say that a
dynamical theory is satisfactory from his point of view. This lack of a
definition of a Machian theory was compounded by the fact that Einstein made no
attempt to implement Mach's ideas directly in GR. In fact,
 in 1902 Poincar\'e \cite{Poincar'e1902, Poincar'e1905} had made an insightful
analysis of Newtonian dynamics, from which a clear defect of it emerged.
The implementation of Mach's principle requires a formulation of dynamics that
eliminates this defect. In this section, I shall give the essence of
Poincar\'e's analysis in a form suitable for the further discussion. A
fuller account can be found in \cite{Barbour2010}.

Consider the dynamics of $N, N\ge 3,$ point particles in
Euclidean space that form a closed dynamical system taken to model the universe.
Only the
distances $r_{ab}, a,b=1,2, ... , N,$ between them are observable. Suppose the
$r_{ab}$ are determined at two instants 1 and 2; the two resulting
sets of $r_{ab}$ define two relative configurations. In Cartesian frames of
reference, the particles will have coordinates $\textbf{x}_a^1$ and
$\textbf{x}_a^2$.  Newton's problem in
laying the foundations of dynamics was this: how can one define the
displacements and hence velocities of the individual particles between the two
instants? The difficulty is that the
Cartesian frames chosen for the two relative configurations are unrelated.
There is nothing intrinsic to the observable $r_{ab}^1$ and $r_{ab}^2$ that
establishes a connection between the frame-dependent $\textbf{x}_a^1$ and
$\textbf{x}_a^2$. This indeterminacy and its consequences for 3-body Newtonian
dynamics are illustrated in Fig.~1.

\begin{figure}
 \begin{center}
 \includegraphics[width=9.5cm]{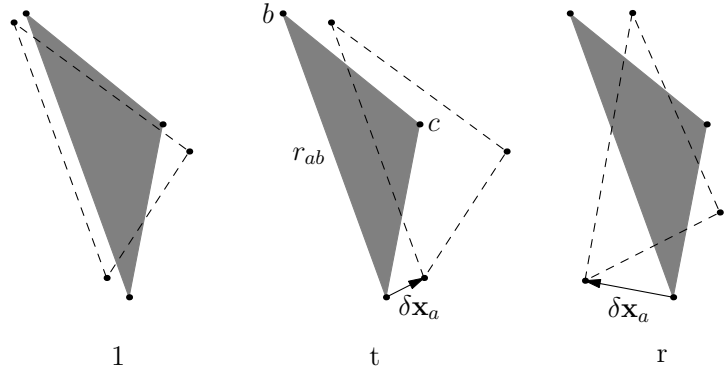}

\caption{The observable separations $r_{ab}^1$ and $r_{ab}^2=r_{ab}^1+\delta
r_{ab}$ define two slighly different (shaded and dashed) triangles and, in the
limit $\delta t\rightarrow 0$, observable initial data $r_{ab}, \dot r_{ab}.$
If, with Newton, one assumes that position in space has physical
meaning, the triangles can be placed separately anywhere in space (in this
representation, on the paper or screen carrying the images). This is equivalent
to choosing arbitrary Cartesian frames of reference for each of them. For
all such choices, $r_{ab}, \dot r_{ab}$ are the same (invariant), but the
Newtonian displacements in space are different. Most importantly, the initial
velocities $\dot\bf{x}_a$, obtained in the limit $\delta t\rightarrow 0$ from
$\delta\bf{x}_a/\delta t$, can be changed from the possibility 1 by the
mere group operations of translating (t) and rotating (d) one
triangle relative to the other. They generate different Newtonian
initial velocities $ \dot\bf{x}_a$, but the $ \dot r_{ab}$ are invariant. This
has the consequence that
different Newtonian evolutions of the observable $r_{ab}$ can arise from
observationally identical initial $r_{ab}, \dot r_{ab}$. Poincar\'e identified
this defect of Newtonian dynamics. Mach's intuition about
dynamics will be realized if this defect is eliminated: \emph{the observable
initial data must determine the observable evolution.}}

 \end{center}

\end{figure}

Newton analyzed the problem of defining individual particle displacements given
only relative positions in an unpublished paper called \emph{De gravitatione}
(see \cite{Barbour2001}, p. 609ff). The associated difficulties led him to
introduce the notions of absolute space and time. However, this had the
consequence that Newtonian dynamics is not as predictive as one might expect,
for it allows many different evolutions of the observable $r_{ab}$ to arise from
identical observable $r_{ab},\dot r_{ab}$. The change in the initial
displacements associated with translations (the difference between 1 and t in
Fig.~1) has no effect on the evolution of the $r_{ab}$ because
of Galilean relativity. However, rotations have a significant
effect because they change the angular momentum in the system. We have the
undesirable consequence that different evolutions can arise from identical
\emph{observable} initial data. One can regard the different relative placings
in Fig.~1 as gauge transformations that leave the observable initial data 
 $r_{ab},\dot r_{ab}$ invariant but not the evolution of the $r_{ab}$.

Poincar\'e found this state of affairs `repugnant' but faced with the manifest
presence of angular momentum in the solar system concluded regretfully that
there was no alternative but to accept the active participation of an invisible
agent in local dynamics. Curiously, he did not consider Mach's suggestion
\cite{Mach1883} (cf. footnote 2) that the universe as a whole might determine
the local inertial
frames in which nearby bodies are observed to evolve.

I shall now explain the simple way to construct
theories that are free from the defect Poincar\'e found in Newtonian dynamics.
As I have written extensively about the definition of
Mach's principle in \cite{Barbour2010} and its implementation in
\cite{Barbour2011}, I will merely give the basic idea and an indication of why
it applies universally, including the theory of dynamical geometry and thus
general relativity (GR). This will show that Newtonian dynamics and GR are very
different precisely because the latter implements Mach's principle in a
formulation that I feel confident Mach would have accepted. Since
absolute time is as invisible as absolute space, I shall replace the derivatives
$\dot r_{ab}$ wrt time by derivatives $r_{ab}'$ wrt an arbitary parameter
$\lambda$.

\section{Best Matching Implements Mach's Principle}

The defect that Poincar\'e identified in Newtonian dynamics has a purely
group-theoretical origin: the generators of Euclidean rotations
change the Newtonian initial data without changing the
observable initial data. The best-matching implementation of Mach's principle
employs the very same generators to solve the problem in Machian dynamics that
they create in Newtonian dynamics. The first thing is \emph{to change
the aim of dynamics}: not to find a
law that each body in the universe satisifies separately in space (as in
Newton's law of inertia) but one that governs the changes of the observable
separations $r_{ab}$ within the universe, treated as a single closed system.
Crucially, an observable initial state $r_{ab}, r_{ab}'$ of the universe
must uniquely determine its observable evolution.

The idea, in its simplest form,\footnote{The extension to
scale-invariant  particle dynamics and dynamical geometry together with
 issues related to time and the expansion of the universe are discussed in
\cite{Barbour2010, Barbour2011}.} is to define a metric on the space $\mathcal
R$ of possible relative configurations of the universe. If this can be done,
then geodesics with respect to this metric will define Machian evolutions in
$\mathcal R$. This is because a
point and direction, which correspond to the observable initial data $r_{ab},
r_{ab}'$ in $\mathcal R$, define a geodesic uniquely. Poincar\'e's requirement
 will be met. Figure 2 illustrates the mechanism that implements this by
creating a metric
on $\mathcal R$ for the case of the 3-body problem.

\begin{figure}
 \begin{center}
 \includegraphics[width=7cm]{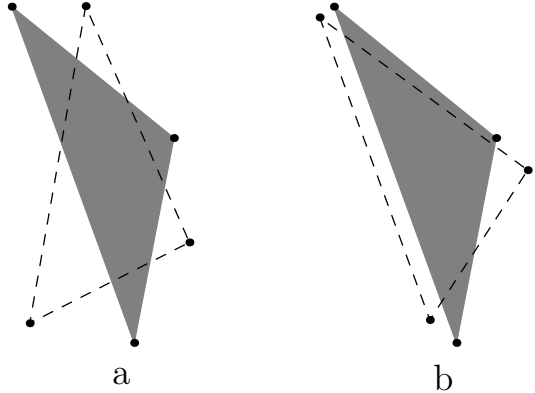}

\caption{a) An arbitrary placing of the dashed triangle
relative to the undashed triangle; b) the best-matched placing reached
by translational and rotational minimization of (\ref{trial}). Best matching
minimizes the expression (\ref{trial}), brings the centres of mass to
coincidence and reduces the net rotation to zero. Note that in this toy
`island-universe' model best matching automatically forces us to take
into account the relative changes within the complete universe.}

 \end{center}

\end{figure}

The mechanism is very simple. To define a metric on $\mathcal R$, we need to
define a
`difference' between any two nearly identical relative configurations of the
system. These are represented by triangles in the 3-body problem. To
measure their difference, we represent one -- either will do -- in an arbitrary
Cartesian frame. The particles then acquire coordinates
$\textbf{x}_a,\,a=1,2,3.$ We then put the other triangle in an initial arbitrary
trial placing somewhere near the first triangle. This corresponds to choosing
some Cartesian frame for the second triangle. Its particles then have
coordinates $\textbf{x}_a+\delta\textbf{x}_a$. In this trial position, we
calculate
\be
\delta s_{trial}:=\sqrt{(E-V)\sum_a
{m_a\over 2}\delta\textbf{x}_a\cdot\delta\textbf{x}_a},\label{trial}
\ee
where $E$ is a constant and $V$ is a function on $\mathcal R$, i.e.,
$V=V(r_{ab})$. It is clear that 
in a Machian approach (\ref{trial}) can have
no physical significance since the trial placing is arbitrary. However, we can
use the generators of translations and rotations to move the second triangle
into the unique position in which
(\ref{trial}) is minimized. This is the \emph{best-matched} position; it is
determined by the triangles alone. The
corresponding extremal value of (\ref{trial}) defines the `difference', or
`distance', between the two relative configurations. This is all that we need to
implement Mach's principle. Note that the best-matched
displacements $\delta\textbf{x}_a^{bm}$ are not defined relative to any
space but relative to the triangle taken to represent the initial configuration.
The role of space as a frame of reference is eliminated. We now consider the
consequences.

As is explained in more detail in \cite{Barbour2011}, the observable
separations $r_{ab}$ in a dynamically closed (island) Machian universe evolve
exactly as in Newtonian theory with two important
differences. First, the total momentum $\textbf{P}$ and angular momentum
$\textbf{L}$ of the universe must be exactly zero and its total energy must 
be $E$. These restrictions appear as gauge-type
\emph{constraints} on the initial data. The vanishing of $\textbf{P}$ is not
significant on account of Galilean relativity, but the constraint $\textbf{L}=0$
eliminates the defect in Newtonian dynamics identified by Poincar\'e. Second, if
the successive relative configurations are placed relative to the initial
configuration in the best-matched
position, the evolution unfolds exactly as in a Newtonian inertial frame of
reference, but this is not introduced prior to the formulation of the
dynamical law. It is emergent. It is important that
subsystems of the universe have their inertial frames determined by the best
matching applied to the universe but can have
nonvanishing angular momentum; the total angular momentum of the universe must
however be zero. The manner in which Newtonian time and the energies of
individual subsystems emerge out
of an initially timeless kinematic framework is described in \cite{Barbour2011}.

\section{General Relativity and Gauge Theory}

Because best matching is based on transformations generated by continuous
groups,\footnote{Lie groups in the case of finite-dimensional systems, their
infinite-dimensional generalizations in the case of gauge theory and dynamical
geometry.} it is a universal mechanism that can be employed to implement
Machian dynamics whenever the configurations of a dynamical system are
invariant under the action of a Lie group. Just as Euclidean transformations
leave the interparticle separations invariant in the
Newtonian $N$-body problem, three-dimensional diffeomorphisms leave
invariant the geodesic distances between points on a 3-manifold on which a
Riemannian 3-metric is defined. As was first shown in \cite{Barbour1982}, this
makes it possible to implement a refined form of best matching between
3-metrics and recover general relativity as a dynamical theory of the
evolution of Riemannian 3-geometry from very basic Machian first principles. 

The key result that establishes the Machian nature of GR is
that the best matching leads directly to the constraint $p^{ij}_{~;j}=0$, where
$p^{ij}$ is the momentum canonically conjugate to the 3-metric $g_{ij}$, and the
semicolon denotes covariant differentiation using the Levi-Civita connection of
$g_{ij}$. This constraint is none other than the well-known momentum constraint
obtained by Arnowitt, Deser and Misner in their 3+1 dynamical analysis of
GR. In fact, best matching explains the presence of
constraints linear in the canonical momenta in both GR and gauge theory
\cite{Barbour2011}.\footnote{The Hamiltonian
constraints quadratic in
the canonical momenta that appear in GR arise because there is
no external absolute time in general relativity.} 

In general dynamical theory, there is a great difference between theories in
which
constraints are present (gauge theory and GR) and theories in
which no constraints are present (Newtonian theory). This difference
reflects the implementation of Mach's principle in GR (and a
related form of it in gauge theory). More detailed discussions that take into
account the treatment of time, scale invariance implemented through conformal
best matching, and a proper constraint analysis can be found
in \cite{Barbour2010, Barbour2011, Gomes2011}.

In summary and to repeat the point made at the start: one cannot
judge whether a theory implements Mach's principle by considering the motion of
a single particle in an inertial frame of reference, be it global or local.
Instead one must ask: What determines the inertial frame of reference? The
answer to this question shows that general relativity is Machian.

\section{Comment on Mashhoon and Wesson's Response}

Mashhoon and Wesson have set out a short response
``Mach, the Universe, and Foundations of Mechanics''
to my comments. They grant that I have given a
possible definition and implementation of Mach's principle in the case
of a spatially closed universe but ``prefer to rely on the
judgement of observational cosmology''. I agree that ultimately
observations must judge the viability of a conceptual scheme and that we
cannot yet survey the whole universe in the way needed to confirm
the correctness of my proposal. However, my concern is with the
foundations of mechanics and the ideal Mach intuited. Let me
make two points.

1. In their original paper, Mashhoon and Wesson stated \emph{as a fact} that
in GR ``the origin of inertia\footnote{When invoking the
universe to explain inertia, Mach was solely concerned with inertia in the sense
of Newton's first law. In accordance with his definition of \emph{inertial
mass} \cite{Mach1960}, p.
264--271, that is an intrinsic property of each individual body. Einstein
introduced much confusion by failing to distinguish the two
meanings of inertia.} remains essentially the same as in Newtonian physics''. My
concern is to counter this claim and to show that Mach's ideas deal with the
most basic foundational question in dynamics: how is motion to be defined? In
attempting to answer this question, I do not think the theoretician should be
restricted to the analysis of direct observations. Indeed, modern science began
when the Greeks introduced theoretical notions into astronomy that went well
beyond the observations \cite{Barbour2001}.

2. Mach recognized (footnote 2) that his ideas involved the entire universe. In
dynamical geometry, this does indeed require the universe to be
spatially closed as Wheeler, following Einstein, advocated (see Isenberg's
article in \cite{Barbour1995}). However, even when boundary conditions
are imposed, the dynamics up to them is best matched and `as Machian as it can
be'. Moreover, my definition of Mach's principle
and its best-matching implementation lead directly to the framework that
Wheeler merely ``let the mathematics tell us'' to adopt (\emph{ibid}, p. 194).

\bibliography{/home/julian/latex/bibtex/julian_barbour_references}
\end{document}